# Machine Learning Accelerated Computational Surface-Specific Vibrational Spectroscopy Reveals Oxidation Level of Graphene in Contact with Water

Xianglong Du[1], Jun Cheng[1,3,4*], Fujie Tang[2,3,4*]

1. State Key Laboratory of Physical Chemistry of Solid Surfaces, iChEM, College of Chemistry and Chemical Engineering, Discipline of Intelligent Instrument and Equipment, Xiamen University, Xiamen 361005, China
2. Pen-Tung Sah Institute of Micro-Nano Science and Technology, Xiamen University, Xiamen 361005, China
3. Laboratory of AI for Electrochemistry (AI4EC), Tan Kah Kee Innovation Laboratory (IKKEM), Xiamen 361005, China
4. Institute of Artificial Intelligence, Xiamen University, Xiamen 361005, China

*Author to whom correspondence should be addressed: chengjun@xmu.edu.cn; tangfujie@xmu.edu.cn

## Abstract

Precise characterization of the graphene/water interface has been hindered by experimental inconsistencies and limited molecular-level access to interfacial structures. In this work, we present a novel integrated computational approach that combines machine-learning-driven molecular dynamics simulations with first-principles vibrational spectroscopy calculations to reveal how graphene oxidation alters interfacial water structure. Our simulations demonstrate that pristine graphene leaves the hydrogen-bond network of interfacial water largely unperturbed, whereas graphene oxide (GO) with surface hydroxyls induces a pronounced ≈100 cm$^{-1}$ redshift of the free-OH vibrational band and a dramatic reduction in its amplitude. These spectral shifts in the computed surface-specific sum-frequency generation spectrum serve as sensitive molecular markers of the GO oxidation level, reconciling previously conflicting experimental observations. By providing a quantitative spectroscopic fingerprint of GO oxidation, our findings have broad implications for catalysis and electrochemistry, where the structuring of interfacial water is critical to performance.

**Introduction**

Graphene is a single-atom-layer two-dimensional material[1,2]. Because the graphene sheet shows the electric conductivity, it has been considered as a promising material for energy conversion and storage[3,4], electrocatalysis[5,6], and chemical sensing[7,8]. Conveniently, the physical and opto-electric property of graphene can be tuned by controlling the oxidation level of graphene. For example, the presence of oxygen-containing functional groups allow graphene oxide (GO) to function as a green oxidant or solid acid[9]; the defects of GO like nanovoids and vacancies induced during its preparation endow it with unpaired spins, which can help to activate small molecules by spin flip process[10]. When GO comes into contact with water or is used to fabricate electrodes, it exhibits characteristics distinct from other carbon-based materials commonly employed in electrochemistry.[11] Unlike traditional carbon materials, GO features a two-dimensional layered structure with a high surface area and a rich abundance of oxygen-containing functional groups[11], such as hydroxyl and epoxide groups. These unique properties enable the functionalization of GO-based electrodes through both covalent and noncovalent strategies, allowing for the fine-tuning of their structural architecture and intrinsic electrochemical properties, when contacting with electrolytes.

However, obtaining molecule level insight into the graphene/GO-water interface is experimentally challenging, because the signal arising from the interface is masked by the bulk contribution. Surface-specific vibrational spectroscopy such as heterodyne-detected sum-frequency generation (HD-SFG), which is a non-linear second order optical process, offer one of the few viable approaches for probing interfacial molecular structures[12,13]. SFG spectroscopy provides unique sensitivity to both the orientation and hydrogen-bonding (H-bonding) environment of interfacial water molecules[14–17]. SFG signals are non-zero only when the centro-symmetry is broken, these from the bulk water are excluded due to the SFG selection rule[14,18] Thanks to its optical transparency in the infrared and visible regions, the suspended graphene-water interface, which is free from the impact of substrate, is especially well-suited for SFG studies, enabling direct observation of molecular-scale phenomena. With this motivation, the researchers explored the pristine graphene-water interface via SFG experimentally (Benderskii[19], Tian[20], Salmeron[21], Nagata[22]) and theoretically (Nagata[22,23], Laage[24,25], and Car[26]), while due to the complexity of the organization of the GO, the GO/water interface has not been explored experimentally. The ultrathin nature of graphene and GO, combined with its intrinsic material properties, poses significant experimental challenges to reveal the molecular structures when contacting with water[27–29].

Moreover, variations in experimental conditions and graphene preparation can introduce additional uncertainties. These challenges are not unique to graphene systems; they are representative of broader difficulties in probing aqueous interfaces, such as those found at electrode/electrolyte[30], nanoconfined[31], or membrane environments[32]. Defects in the graphene sheet may lead to GO during experiments, resulting in the formation of hydroxyl groups that can alter the SFG signal[33]. Given these experimental uncertainties, in this context, computational spectroscopic techniques provide powerful capabilities for precise modeling and molecular-level insight[34–36]. These methods allow for the construction of models from first principles and enable the direct calculation of spectral responses under varying interfacial configurations. As such, they not only help to identify the spectral origins with high accuracy but also serve as a benchmark for interpreting and validating experimental measurements.

In this work, we compute the SFG spectra of suspended graphene–water and suspended GO–water interfaces using machine learning molecular dynamics (MLMD) simulations to uncover their molecular structures and corresponding spectral responses. We propose that the observed ~100 cm$^{-1}$ redshift and reduced peak intensity are likely attributable to hydroxyl OH groups on the free side of the GO sheet, which can serve as spectral signatures of the GO. Through a comprehensive analysis, we find that the H-bonds formed between the chemical absorbed groups and interfacial water molecules greatly reduced the SFG amplitude of the negative H-bond peak, leading to the observed reduced amplitude in the experiment. Moreover, we demonstrate that pristine graphene produces only a weak SFG signal, consistent with the reported simulation results[23–26,37]. The observed SFG response primarily originates from the topmost water layers in contact with the suspended pristine graphene. This study highlights the essential role of computational spectroscopic methods in accurately assigning and interpreting vibrational spectra at complex interfaces.

## Methods

**Construction of Initial Structures.** We first construct graphene/water and GO/water interfaces with varying oxidation levels—12.5%, 25.0%, and 50.0%—using the PackMol package[38]. The schematics of graphene/water and GO/water interfaces are shown in Fig. 1(a) and 1(b), respectively. The number of carbon atoms and functional groups in each system are as follows: for the 12.5% oxidation level, there are 144 carbon atoms, 10 epoxide groups, and 8 hydroxyl groups; for the 25.0% oxidation level, 144 carbon atoms, 18 epoxide groups, and 18 hydroxyl groups; and for the 50.0% oxidation level, 144 carbon atoms, 36 epoxide groups, and 36 hydroxyl groups.

**Molecular Dynamics Simulation.** We performed the MLMD simulations at the air/water interface, graphene/water interfaces and GO/water interfaces with 12.5%, 25% and 50% oxidation levels, respectively. For the machine learning model, we used the Deep Potential (DP) model with message passing as previous study[39] had shown its better performance in the charged system. The detailed comparison with DeepMD with short range descriptor is documented in the SI. The training data set of potential energy and force was generated by using the CP2K package[40]. All density functional theory (DFT) calculations were conducted with the CP2K/QUICKSTEP module, which uses a mixed Gaussian and plane-wave basis set. The revised Perdew-Burke-Ernzerhof (revPBE) functional was applied to describe exchange-correlation energies[41,42], with Grimme's D3 dispersion correction included in all calculations[43]. A triple-zeta valence plus two polarization (TZV2P-MOLOPT-PBE-GTH) was used[44,45], with a plane-wave cutoff of 800 Ry. Core electrons were represented by Goedecker-Teter-Hutter (GTH) pseudopotentials[46], and the convergence threshold of self-consistent field optimization was set to $1 \times 10^{-6}$ a.u.. For the initial structures of MD simulations, the cell sizes of air/water and graphene/water interfaces were $29.52 \times 25.566 \times 70.000$ Å, while the cell sizes of GO/water interfaces were $29.712 \times 25.756 \times 70.000$ Å, $29.794 \times 26.014 \times 70.000$ Å and $29.998 \times 25.850 \times 70.000$ Å for 12.5%, 25% and 50% oxidation levels, respectively. The training dataset for the DP model was generated using the ai2-kit[47] workflow, which is similar to the DP-GEN workflow[48]. For details on the DP model and ai2-kit training parameters, please refer to the SI.

Once the DP models were obtained, we performed machine learning molecular dynamics (MLMD) simulations to sample the molecular configurations. The initial structures used in the MLMD simulations are as follows: air/water interface (752 water molecules), graphene/water interface (576 carbon atoms, 752 water molecules), 12.5% GO/water interface (576 carbon atoms, 752 water molecules, 32 hydroxyl groups and 40 epoxide groups), 25.0% GO/water interface (576 carbon atoms, 752 water molecules, 72 hydroxyl groups and 72 epoxide groups), and 50.0% GO/water interface (576 carbon atoms, 752 water molecules, 144 hydroxyl groups and 144 epoxide groups). All MLMD simulations were carried out in the *NVT* ensemble using the JAX-MD code at 300 K for 500 ps[49], producing trajectories for spectral calculations and structural analysis.

**SFG Calculations.** We calculated the SFG spectra using the surface-specific velocity-velocity autocorrelation function[50] with the coordinates and velocities of water molecules and hydroxyl groups. This method enables one to compute the SFG spectra with a reasonable s/n ratio solely from the MD trajectories with few hundred picoseconds. Within this formalism, the resonant part of the SFG susceptibility, $\chi_{xxz}^{(2),R}(\omega)$, can be given as:

$$\chi_{xxz}^{(2),R}(\omega) = \frac{Q(\omega)\mu'(\omega)\alpha'(\omega)}{i\omega^2}\chi_{xxz}^{ssVVAF}(\omega), \quad (1)$$

$$\chi_{xxz}^{ssVVAF}(\omega) = \int_0^\infty dt e^{-i\omega t} \langle \sum_i g_{ds}(z_i(0))\dot{r}_{z,i}^{OH}(0)\frac{\dot{\vec{r}}_i^{OH}(t)\cdot\vec{r}_i^{OH}(t)}{|\vec{r}_i^{OH}(t)|}\rangle, \quad (2)$$

where $g_{ds}(z_i)$ is the truncation function for the dividing surface to selectively extract the vibrational responses of water molecules near the interface.

$$g_{ds}(z_i) = \begin{cases} 0 \text{ for } z_i \geq z_{ds} \\ 1 \text{ for } z_i < z_{ds} \end{cases}, \quad (3)$$

where $z_{\text{ds}}$ is the z-coordinate of the dividing surface and $z_i$ is the z-coordinate of the O atom of the *i*th O-H bond. The $z_{\text{ds}}$ value was set to decouple the responses of the GO/water interface and graphene/water interface from the water/air interface. We set the origin point as the averaged position of C atoms for the graphene/water interface. The $z_{\text{ds}}$ value is set to 11 Å for the graphene/water and GO/water system.

The frequency-dependent induced transition dipole moment and polarizability due to the solvation effects were included by using the frequency-dependent transition dipole moment ($\mu'(\omega)$) and polarizability ($\alpha'(\omega)$)[51,52]:

$$\mu'(\omega) \equiv \left(1.377 + \frac{53.03(3737.0 - \omega)}{6932.2}\right)\mu^0, \tag{4}$$

$$\alpha'(\omega) \equiv \left(1.271 + \frac{5.287(3737.0 - \omega)}{6932.2}\right)\alpha^0, \tag{5}$$

where $\omega$ is in cm$^{-1}$. $\mu^0$ and $\alpha^0$ are permanent dipole moments and permanent polarizability of OH chromophores, respectively. $Q(\omega)$ is the quantum correction factor given by[53]:

$$Q(\omega) = \frac{\beta\hbar\omega}{1 - \exp(-\beta\hbar\omega)}, \tag{6}$$

where $\beta = 1/k_BT$ is the inverse temperature.

## Results and Discussion

**Theoretical SFG Spectra of Suspended GO/Water and Graphene/Water Interfaces**

First, we show the calculated SFG $\text{Im}\chi^{(2)}_{ssp}$ spectra of water at the suspended graphene/water interface, the suspended GO/water interface as well as the water/air interface in Fig. 1(c), respectively. As one can see, our graphene/water SFG spectrum exhibits a negative ~3400 cm$^{-1}$ broad peak and a high-frequency positive 3700 cm$^{-1}$ peak. This is consistent with previous theoretical works.[23,26,37] The negative peak originates from the H-bonded O–H stretching mode of water molecules at the suspended graphene/water interface, with the dipole orientation pointing toward the bulk water. The amplitude and frequency of this negative peak are nearly identical to those observed at the water/air interface, indicating that the pristine graphene sheet induces only minor modifications to the H-bond strength. The high-frequency peak at the pristine graphene/water interface exhibits a slight red shift relative to that at the water/air interface, with the same sign and similar amplitude, but a frequency difference of approximately $\Delta f \approx 40$ cm$^{-1}$. Interestingly, as shown in the Fig.1(c), in the calculated SFG spectra of the suspended GO/water interface at an oxidation level of 12.5%, a pronounced red-shifted high-frequency peak with reduced amplitude emerges at approximately $f \approx 3625$ cm$^{-1}$, along with a weak shoulder near $f \approx 3690$ cm$^{-1}$. Compared to the high-frequency peak observed at the water/air interface, this corresponds to a frequency shift of approximately $\Delta f \approx 120$ cm$^{-1}$.

Notably, two recent SFG studies on the suspended graphene–water interface have reported substantial discrepancies in their findings, underscoring the complexity and sensitivity of such measurements. Bonn et al.[22] reported that the SFG spectral feature of the suspended graphene/water interface is similar with those of the water–air interface. Both systems display a prominent negative peak near 3400 cm$^{-1}$, corresponding to H-bonded OH groups pointing toward the bulk, and a positive peak around 3650 cm$^{-1}$, associated with free OH groups oriented toward the interface— toward air in the water/air case and toward the graphene in the suspended graphene/water case. The main difference lies in the position of the free OH peak: the suspended graphene/water interface shows a redshift of approximately 30 ± 10 cm$^{-1}$ relative to the water/air interface, suggesting weak interactions between the graphene surface and interfacial water. In contrast, Tian et al.[20] reported a significantly larger redshift of ~100 cm⁻¹ and a roughly 50% reduction in the amplitude of the SFG signal compared to the water–air interface. These observations point to much stronger interactions between graphene and interfacial water in their study[20], highlighting the sensitivity of interfacial water structure to subtle variations in experimental conditions or graphene preparation.

The differences in peak frequency and amplitude between the GO/water interface (Δf ≈ 120 cm⁻¹, with an amplitude comparable to that of the water/air interface) and the pristine graphene/water interface (Δf ≈ 40 cm⁻¹, with approximately half the amplitude of the water/air interface), as predicted by computational spectroscopic methods, closely align with the discrepancies observed in the experimental results. In fact, surface charges or functional groups attached on the GO can significantly influence interfacial vibrational spectra. For instance, Paesani and coworkers[37] modeled charged graphene–water interface using the MB-pol data-driven many-body potential. Their results demonstrated that the spectral response changes systematically with surface charge; Kumar and collaborators[54,55] reported SFG spectra for both graphene/water and GO/water interfaces, while their calculated pristine graphene/water spectrum that differs substantially from all existing experimental data. Nevertheless, these results highlight the impact of the surface functional groups on the modifications of spectral shape.

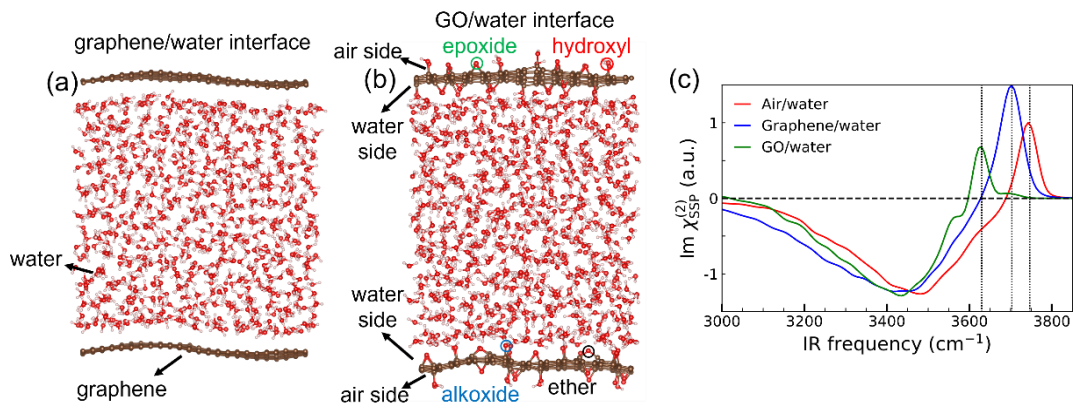

**Fig. 1.** The schematic of the graphene/water interface (a) and the GO/water interface (b), adopted from MLMD simulation trajectory. Note that the corresponding geometry is suspended graphene/GO is on the top of water. (c) The calculated $\mathrm{Im}\chi_{ssp}^{(2)}$ spectra of the air/water interface (red), suspended graphene/water interface (blue) and suspended GO/water interface (green). The suspended GO is at 12.5% oxidation level. The dashed line serves as the zero lines.

**Role of Functional Groups on the Graphene Oxide in Modulating the SFG Response**

When in contact with water, the functional groups on the GO sheet can form H-bonds with interfacial water molecules, thereby altering the SFG spectral response. Before interpreting the spectral changes observed in Fig. 1(c), we first examine the distinct nature of the functional groups present on the GO sheet. In GO, carbon atoms covalently bonded to oxygen-containing functional groups—such as hydroxyl, epoxide groups—are *sp³* hybridized, shown in Fig. 2(a). The epoxide group can interact with interfacial water molecules, furthermore transforms into alkoxide or ether group. These functional groups attached regions are considered oxidized and disrupt the extended *sp²*-conjugated network characteristic of the original honeycomb lattice of pristine graphene. In contrast, the unmodified *sp²* domains represent the unoxidized regions. The *sp³*-hybridized carbon clusters are uniformly distributed but exhibit slight, random displacements above or below the graphene plane, as demonstrated in Fig. 1(b).

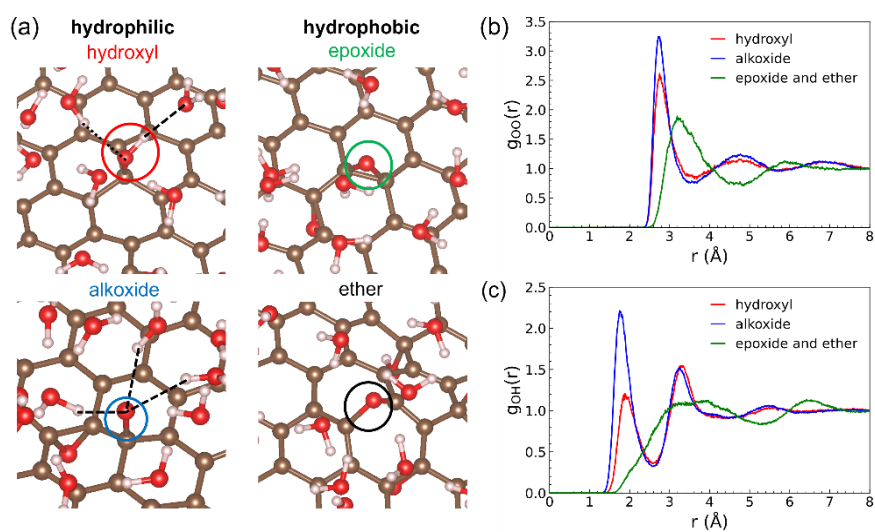

**Fig. 2.** (a) The schematic illustrates the functional groups at the GO/water interface. The hydrophilic groups include hydroxyl (red) and alkoxide (blue), while the hydrophobic groups include epoxide (green) and ether (black). The radial distribution functions $g_{OO}(r)$ (b) and $g_{OH}(r)$ (c) of the functional groups and water molecules, respectively. The red line represents the distribution between hydroxyl groups and interfacial water molecules, the blue line corresponds to alkoxide–water interactions, and the green line represents the interactions between epoxide/ether groups and interfacial water molecules.

Moreover, among these functional groups, the hydroxyl and alkoxide groups exhibit hydrophilic characteristics and can act as H-bond donors or acceptors with interfacial water molecules. In contrast, the slightly non-polar epoxide and ether groups display hydrophobic behavior, disrupting the H-bonding network near the GO surface. In Fig. 2(b) and 2(c), we present the calculated radial distribution functions, $g_{OO}(r)$ and $g_{OH}(r)$, between the functional groups and water molecules. As expected, the peaks of $g_{OO}$(r) at around $r$ = ~2.7 Å and $g_{OH}$(r) at around $r$ = ~1.8 Å for the hydroxyl and alkoxide groups are characteristic of H-bonding, similar to those observed in liquid water[56,57]. This indicates that hydroxyl and alkoxide groups can form hydrogen bonds with strengths comparable to those in bulk water. In contrast, the larger oxygen–oxygen ($r$ > ~3.3 Å) and oxygen–hydrogen ($r$ > ~3.0 Å) distances associated with the epoxide and ether groups suggest that H-bonding between these groups and interfacial water is negligible, consistent with our previous interpretations.

After discussing the impact of functional groups on the molecular structure of interfacial water near GO, we now turn our attention to their influence on the SFG spectral response of the GO/water interface. To this end, we decompose the SFG spectrum of the GO/water interface into two components: one corresponding to the O–H stretching modes associated with hydroxyl groups on the GO sheet, and the other to the spectral contribution from interfacial water, as shown in Fig. 3(c). Since both sides of the GO sheet are in contact with water, we calculated the spectral responses from each surface separately. Interestingly, the hydroxyl groups on the air-facing side give rise to a positive peak at around 3625 cm$^{-1}$, accompanied by a shoulder near 3575 cm$^{-1}$. In contrast, the hydroxyl groups on the water-facing side contribute a broad negative peak centered around 3100 cm$^{-1}$, indicating strong H-bonding with interfacial water molecules. The SFG spectrum of the interfacial water itself exhibits a characteristic "positive–negative–positive" pattern, with a broad negative peak around 3450 cm$^{-1}$. Notably, the two positive features near 3100 cm$^{-1}$ and 3700 cm$^{-1}$ correspond to interfacial water molecules involved in strong H-bonding with surface functional groups and to the free O–H stretch, respectively—the latter being consistent with the positive peak observed at the graphene/water interface.[22,23]

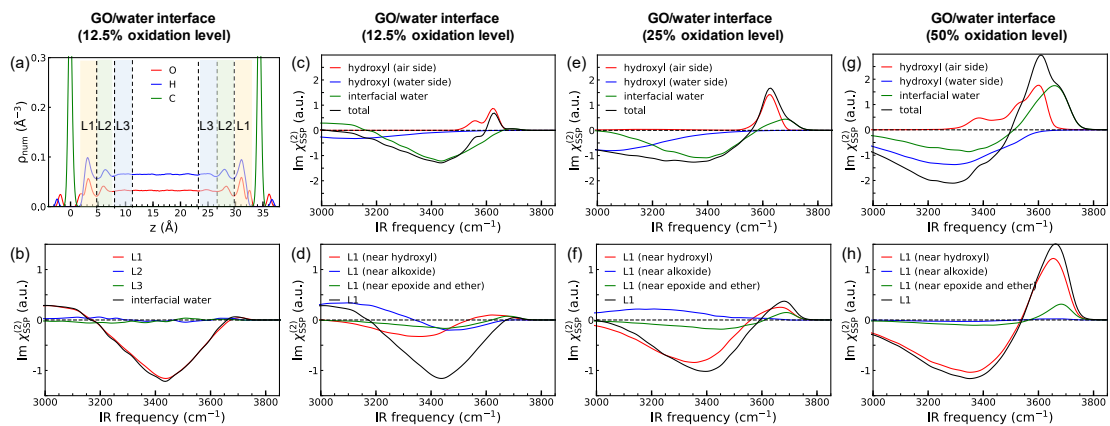

**Fig. 3.** (a) The number density profiles of the oxygen, hydrogen, and carbon atoms along the surface normal calculated by using MLMD simulation trajectory. (b) The decomposed $\mathrm{Im}\chi^{(2)}_{ssp}$ spectra of the interfacial water from the suspended GO/water interface at 12.5% oxidation level, showing the contributions from different water layers. The red curve represents contributions from the 1$^{st}$ water layer, the blue from the 2$^{nd}$ water layer, and the green from 3$^{rd}$ water layer. (c), (e) and (g) The decomposed $\mathrm{Im}\chi^{(2)}_{ssp}$ spectra of the suspended GO/water interface showing the contributions from hydroxyl groups and interfacial water at 12.5% oxidation level, 25% oxidation level and 50% oxidation level, respectively. The red curve represents contributions from the air side, the blue from the water side, and the green from interfacial water. (d), (f) and (h) Spectral contributions to the SFG spectra from water molecules in the first interfacial layer at 12.5% oxidation level, 25% oxidation level and 50% oxidation level, respectively. The red curve represents contributions from water molecules near hydroxyl groups, the blue curve corresponds to those near alkoxide groups, and the green curve is the sum of the contributions from epoxide and ether groups. The black curve shows the total response from all water molecules in the first layer.

We now focus on the SFG spectral contributions from the interfacial water. First, the layer-dependent spectra of the GO/water interface are presented in Fig. 3(b). As expected, the dominant contribution to the SFG response arises from the first water layer, as the structural heterogeneity in this region is primarily governed by the influence of functional groups attached to the GO sheet. To further investigate these effects, we calculated the SFG spectral contributions from interfacial water near different types of functional groups, as shown in Fig. 3(d). Water molecules near hydroxyl groups tend to donate their OH bonds to the oxygen atom of hydroxyl group, forming a slightly upward-pointing configuration that gives rise to a positive peak around 3600 cm$^{-1}$. In contrast, water molecules near hydrophobic groups—such as epoxide and ether—retain a free OH bond pointing toward the GO surface, resulting in a higher-frequency positive peak around 3700 cm$^{-1}$. Interestingly, water molecules adjacent to alkoxide groups exhibit a distinctive "positive–negative" spectral pattern, with a broad positive peak near 3100 cm$^{-1}$ and a negative peak around 3450 cm$^{-1}$. The former indicates strong H-bonding between alkoxide groups and interfacial water, and its amplitude is comparable to the negative contribution at ~3100 cm$^{-1}$ from hydroxyl groups on the water-facing side—leading to a partial cancellation of these effects. Consequently, the overall SFG spectrum of the GO/water interface features a broad negative peak centered around 3400 cm$^{-1}$ and a positive free-OH-like peak near 3625 cm$^{-1}$. This spectral profile is reminiscent of the graphene/water interface but reflects a more complex interplay of constructive and destructive contributions from various functional groups.

**SFG Spectral Signatures of Oxidation Levels of the GO in Contact with Water**

Our investigation focuses on the role of functional groups on the GO sheet in modulating the SFG spectra at the GO/water interface, revealing a complex cancellation effect arising from the opposing contributions of different groups. As the oxidation level of GO increases, the number of functional groups grows accordingly, leading to significant changes in the intrinsic properties of graphene. To examine this effect, we calculated the SFG spectra of the GO/water interface at varying oxidation levels—12.5%, 25.0%, and 50.0%—as shown in Fig. 3(c), (e), and (g), respectively. Before discussing how the SFG spectra evolve with oxidation level, we first comment on the corresponding structural changes of GO. At low oxidation levels, much of the graphene surface remains unoxidized but becomes locally charged due to the presence of functional groups. These functional groups are randomly distributed, and they attract interfacial water molecules. As the oxidation level increases, the unoxidized regions shrink, while the interaction between water molecules and the increasingly dense functional groups becomes more dominant. These two effects—preservation of unoxidized graphene and increased functionalization—compete with each other. At the highest oxidation level (e.g., 50%), the interfacial properties are primarily governed by the functional groups, with the influence of the remaining unoxidized areas becoming negligible.

As expected, the SFG spectral changes reflect the underlying structural modifications. First, the SFG amplitude of the hydroxyl groups on the air side continuously increases with rising oxidation levels, as the SFG signal is proportional to the number of chromophores. Since this peak arises solely from surface hydroxyl groups introduced by oxidation and its frequency remains unchanged with the changes of the oxidation levels, it can serve as a spectral signature of the GO oxidation level. For the hydroxyl groups on the water side, the SFG amplitude shows a slight increase, while the peak position gradually shifts to higher frequencies (blue shift). This indicates a weakening of the H-bonds between hydroxyl groups and interfacial water molecules, likely due to steric effects introduced by the increasing density of functional groups on the GO surface. This trend is consistent with the changes observed in the radial distribution functions $g_{OO}(r)$ and $g_{OH}(r)$ between functional groups and water molecules, as shown in Fig. 2(b) and (c). Moreover, the SFG response of interfacial water captures the competition between two effects: the preservation of unoxidized graphene regions and the increasing functionalization of GO. At low oxidation levels, the interfacial water exhibits a characteristic "positive–negative–positive" spectral pattern, with a broad negative peak centered around 3450 cm$^{-1}$. As the oxidation level increases, this transforms into a "negative–positive" feature. When the oxidation reaches 50%, the amplitude of this "negative–positive" pattern further increases, indicating the dominance of functional group–induced structural changes at the interface.

This interpretation is further supported by the decomposed spectra of the first water layer near the GO sheet, shown in Fig. 3(d), (f), and (h). At low oxidation levels, the spectral contributions from water near all functional groups are less prominent than those from the unoxidized graphene regions, as illustrated in Fig. 3(d). Notably, the absence of the free-OH-like peak around 3625 cm$^{-1}$ can be attributed to the slight positive surface charge induced by the functional groups on GO. This observation is consistent with previous simulations of charged graphene–water interfaces using the MB-pol model[37]. In stark contrast, at higher oxidation levels—as shown in Fig. 3(f) and (h)—the SFG response of interfacial water becomes dominated by contributions from water molecules interacting with hydroxyl groups. In both the 25.0% and 50.0% oxidation cases, this interaction leads to pronounced changes in the spectral profile. These findings confirm that the appearance and evolution of the free-OH-like peak, along with overall spectral variations, provide a reliable and sensitive vibrational signature for identifying the oxidation level of GO at aqueous interfaces.

## Conclusion

In this study, we have elucidated the molecular origins of SFG spectral features at suspended graphene and GO-water interfaces using machine learning molecular dynamics combined with computational vibrational spectroscopy. Our results reveal that pristine graphene preserves the H-bond network of interfacial water, whereas the introduction of hydroxyl and epoxide functional groups on GO induces substantial modifications in interfacial structure and vibrational response. In particular, the emergence and evolution of a redshifted free-OH-like SFG peak, along with a significant amplitude reduction in the H-bonded OH region, are shown to correlate directly with the oxidation level of GO. These spectroscopic signatures provide a robust molecular-level descriptor of GO functionalization and can reconcile conflicting experimental observations. More broadly, this work demonstrates how state-of-the-art computational spectroscopy can serve as a quantitative tool to benchmark surface-sensitive measurements and unravel complex interfacial phenomena relevant to electrochemistry, catalysis, and energy applications.

## Associated Content

### Data availability

The data that support the findings shown in the figures are available from the corresponding author upon reasonable request.

### Supporting Information

The Supporting Information is available free of charge at https://pubs.acs.org/xxx. Details of the ML model constructions, training parameters and validations; the density profile and angular distribution of GO/water interface at different oxidation level; the decomposed SFG spectra of water of different layers from the suspended GO/water interface at different oxidation level.

**Notes**

The authors declare no competing financial interest.

**Author contributions**

F.T. and J.C. designed the study. X.L.D. conducted the MLMD simulations. X.L.D. and F.T. performed data analysis. X.L.D., J.C. and F.T. wrote the manuscript. All authors contributed to interpreting the results and refining the manuscript.

## Acknowledgments

F.T. acknowledges the National Key R&D Program of China (Grant No. 2024YFA1210804) and a startup fund at Xiamen University. J.C. acknowledges the National Natural Science Foundation of China (Grant Nos.22021001, 22225302, 21991151, 21991150, 92161113, and 20720220009) and the Laboratory of AI for Electrochemistry (AI4EC) and IKKEM (Grant Nos. RD2023100101 andRD2022070501) for financial support. This work used the computational resources in the IKKEM intelligent computing center.